\documentstyle[prl,aps,psfig,twocolumn]{revtex} 
\begin{document}
\draft
\twocolumn[\hsize\textwidth\columnwidth\hsize\csname
@twocolumnfalse\endcsname
\title{Levy-Nearest-Neighbors Bak-Sneppen Model}
\author{R. Cafiero$^1$, P. De Los Rios$^2$, A. Valleriani$^3$, and
J. L. Vega$^4$}
\address{$^1$ P.M.M.H., Ecole Sup\'erieure de Physique et 
de chimie Industrielles, 10, rue Vaquelin, 75231 Paris, France.}
\address{$^2$ Institut de Physique Theorique, Universit\'e de Fribourg,
CH-1700 Fribourg, Switzerland.}
\address{$^3$ Max-Planck-Institute of Colloids and Interfaces - 
Theory Division Kantstrasse 55, D-14513 Teltow-Seehof, Germany}
\address{$^4$ BANCOTEC GmbH, Calwerstra\ss e 33, D-70173 Stuttgart, Germany}
\date{\today}
\maketitle

\begin{abstract}

We study a random neighbor version of the Bak-Sneppen model,
where "nearest neighbors" are chosen according to a probability
distribution decaying as a power-law of the distance from the active site,
$P(x) \sim |x-x_{ac }|^{-\omega}$. All the exponents characterizing the
self-organized critical state of this model depend
on the exponent $\omega$. As $\omega \to 1$ we recover
the usual random nearest neighbor version of the model. 
The pattern of results obtained for a range of values of $\omega$ is
also compatible with the results of simulations 
of the original BS model in high dimensions. Moreover, our 
results suggest a critical dimension $d_c=6$ for the Bak-Sneppen model, 
in contrast with previous claims.

\end{abstract}
]
\narrowtext

Since its introduction, the Bak-Sneppen (BS)  model~\cite{BS93}
has had much success as perhaps the simplest and yet non-trivial
self-organized critical (SOC) extremal model. 
Understanding its behavior is
therefore very important to get an insight in the 
behavior of other SOC extremal models~\cite{PMB96}.

The BS model is easily defined: To each site $i$ on a hypercubic 
lattice in $d$ dimensions is assigned a random variable $f_i$ 
taken from a probability
distribution $p(f)$, say, uniform in $[0,1]$. 
Then at each time-step the site $i$
with the smallest $f_i$ is chosen (it is called the {\it
active} site), and its variable and the variables
of its $2d$ nearest-neighbors are updated taking them from $p(f)$.
As a result of this dynamics, 
the system organizes in a stationary state where 
almost all the variables $f_i$ are above a threshold $f_c$.
Moreover, in this state, the dynamics of the model
has self-similar features: each update of the minimum variable
triggers a local avalanche of updates; the time
durations of the avalanches obey a power-law distributions
characterized by an exponent $\tau$. The number of sites touched
by an avalanche of duration $t$ grows like $t^\mu$.
Also the first return times (defined as the times between two
 successive returns
of the activity to the same site) and the all return times 
(the times between the first passage of the activity 
on a site and any successive return to 
the same site) are power-law distributed, with exponents
$\tau_f$ and $\tau_a$ respectively.

Not all of the above exponents are independent. Indeed it is possible to 
show, from renewal theory, that $\tau_a + \tau_f = 2$ if $\tau_a <1$,
$\tau_a = \tau_f$ if $\tau_a>1$~\cite{fisher}. 
Recently, a non trivial relation between $\tau$ and $\mu$
has been unveiled in \cite{Maslov96,MDM98} exploiting the 
hierarchical structure
of the update avalanches (each avalanche is made up of
smaller avalanches, and so on down to the microscopic scale). 
Both relations are
satisfied for $d=1$ with $\tau \sim 1.07$ and $\mu \sim 0.42$, 
$\tau_a \sim 0.42$
and $\tau_f \sim 1.58$~\cite{BS93,PMB96}.

\begin{figure}
\centerline{\psfig{file=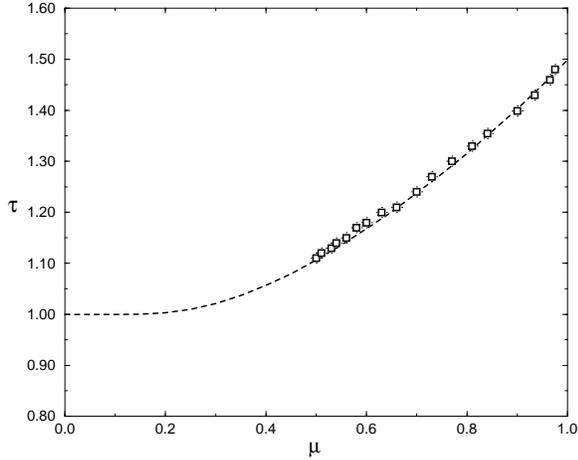,width=9cm,angle=270}}
\caption{Avalanche exponent $\tau$ vs. $\mu$ for different Levy
exponents $\omega$. The dashed line is the exact relation
between the two exponents as from [5]. The data from the simulations
and the exact relation are in excellent agreement. Error bars are comparable
with the symbol dimensions.}
\label{Fig: fig1}
\end{figure}

The only known exactly solved version of the BS model is the Random
Nearest-Neighbor (RNN) model: there "nearest neighbors" are chosen
at random over the lattice\cite{FSB93}. As a result geometric correlations
typical of low dimensions are lost, and the RNN can be considered
as a Mean Field version of the BS model.
The exponents of the RNN model are known to be
$\tau=\tau_a=\tau_f=3/2$ and $\mu=1$\cite{dBDFJW94}.
In particular these exponents satisfy both $\tau_a = \tau_f$
and the relation between $\tau$ and $\mu$ proposed in \cite{Maslov96}.
In \cite{MDM98} this relation has been carefully studied, and it has been
"graphically" explicited (see Fig.\ref{Fig: fig1}).
The knowledge of the two exponent relations
has given confidence in high-dimensional simulations
\cite{DMV98} whose main result is
that the upper critical dimension of the model is $d_{uc}=8$ (the 
upper critical dimension is the dimension
above which the exponents should take the RNN values).
This conclusion is at odds with previous claims, based on analogies 
of the BS model with directed percolation, that $d_{uc}$ should be $4$.
A further result from \cite{DMV98} is the presence of two regimes as the 
dimension $d$ of the system increases:  for $d \le 3$ the model is 
recurrent ($\tau_a<1$; in random walk theory 
recurrence means that every site of the lattice is touched an infinite
number of times with probability one),  whereas for $d \ge 4$ 
the model is transient ($\tau_a>1$; transience means that there is a finite 
probability, smaller than $1$, that a site will be touched by the process), 
yet non trivial ({\it i.e.},
different from the RNN model) as long as $d < 8$.
Actually, before \cite{DMV98}, a further exponent relation 
was believed to hold, namely
$\mu = \tau_a$. In \cite{DMV98} this relation is shown not
to hold for $d>2$. 
Indeed, since $\mu \le 1$ always and $\tau_a(RNN) = 3/2$, it is straightforward
to conclude that at least as soon as $\tau_a > 1$, $\tau_a \ne \mu$.
The dimensionality $d=3$, with $\tau_a <1$ and $\mu \ne \tau_a$,
can be therefore
considered as representative of a further regime within the recurrent one.


The BS model shows therefore an extremely rich behavior changing the
dimensionality $d$ of the system. Yet, high-dimensional simulations 
are always susceptible of strong finite-size corrections, 
and the good convergence of the results is difficult to prove.
In this Communication we propose a new
way to interpolate between the $d=1$ and the RNN models:
The "nearest neighbors" of the active site $x_{ac}$ are chosen at random
over the lattice, but with a probability that is a power-law
decreasing function of the distance from it
\begin{equation}
P(x) \sim |x-x_{ac}|^{-\omega}\;\;.
\label{nn law}
\end{equation}
As we will show, varying $\omega$ we find the same behavioral pattern as
found in \cite{DMV98} varying $d$. We name this model 
the Levy-Random-Nearest-Neighbor
model (LRNN; here the use of the word {\it Levy} is somehow an abuse
since we use also $\omega > 3$).

As a loose analogy, we recall that the same idea has been applied also to the 
$d=1$ Ising model with interactions
decaying as (\ref{nn law}), and indeed it has been found that, varying
$\omega$ it is possible to go from the $d=1$ model to mean-field like results
\cite{Cannas95}. 

Simulations are performed over $1d$ lattices of up to $2^{19}$ sites,
with growing sizes showing stability of the exponents.

In Fig.\ref{Fig: fig1} we show the $\tau$ avalanche exponents
for different values of $\omega$ plotted against the corresponding values
of $\mu$. All the $\mu/\tau$ pairs nicely satisfy the exponent relation between
the two exponents obtained in \cite{DMV98}. 
This is a first important check of the consistency of our 
simulations and of the
exponent relation.

\begin{figure}
\centerline{\psfig{file=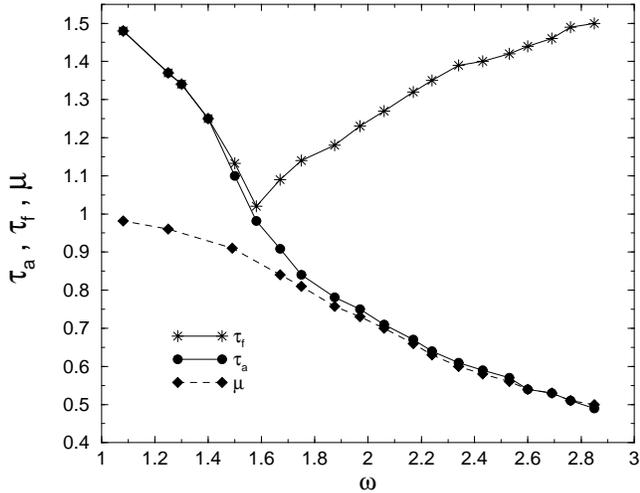,width=8.5cm,angle=0}}
\caption{Values of the different relevant exponents, $\mu$, $\tau_a$ and
$\tau_f$, as a function of the Levy exponent $\omega$. }
\label{Fig: fig2}
\end{figure}

In Fig.\ref{Fig: fig2} we plot the $\tau_a$, $\tau_f$ and $\mu$ exponents 
for different values of $\omega$ from $\omega=3$ to $\omega=1^+$ (this
extreme value is not shown since simulations become extremely
difficult due to the non normalizability of the distribution
if $\omega=1$). Many important aspects of the model can be discussed
looking at Fig.\ref{Fig: fig2}.
We find that the exponent relation between $\tau_a$ and $\tau_f$ 
is satisfied both when $\tau_a < 1$ and when $\tau_a>1$. 
Therefore this result and Fig.\ref{Fig: fig1}
confirm the validity of the two already known exponent relations.
Moreover, we see that the exponents tend to their RNN values as $\omega \to 1$. 
This result should have been expected. Indeed the probability distribution
(\ref{nn law}) is normalizable in the thermodynamic limit 
only as long as $\omega > 1$; when $\omega < 1$ then the normalization
is ruled by the length of the lattice
\begin{equation}
\int_1^L x^{-\omega} \sim L^{1-\omega}
\label{normalization}
\end{equation}
that diverges in the limit of infinite lattice size $L$. Therefore
the distribution $P(x)$, properly normalized, degenerates to 
$0$, just as in the RNN case,
where the normalization is $1/L$ (case $\omega=0$).


The opposite case, $\omega \to \infty$, is also intriguing. Indeed
we could naively expect the $d=1$ limit to be recovered when
$\omega \ge 3$. In that case the 
average distance of the "neighbors" and its variance are finite, just as for
nearest neighbors. We would thus expect the $\omega > 3$ case to belong to the
same universality class as the original BS model, but this conclusion
is not correct.
In order to shed light on this problem, we also performed simulations
taking the neighbors according to a distribution exponentially decreasing
with respect to the distance from the active site $x_{ac}$. In this case,
instead, we nicely recover the known $d=1$ exponents of the BS model.
We conclude therefore that the presence of diverging moments of order higher 
than two drives the system out of its nearest neighbor fixed
point. The latter holds instead whenever the distribution of the 
random neighbors has all its moments finite.
This result, although non trivial, is not new in (annealed or quenched)
disordered systems. For example, it has been
shown that diverging moments of order higher
than two in the disorder distribution
can change the universality class of directed polymers
in random environments, and of the related Kardar-Parisi-Zhang
surface growth equation\cite{HHZ95}.

It is well known in classical random walk theory that random walks
with a jump probability distribution with finite variance belong 
to the Gaussian universality class. Random walks with
infinite higher moments have a microscopic structure that
is different from a Gaussian one, mainly made of clusters of points
seldom separated by long jumps. On larger and larger length scales, this
cluster structure disappears, and the walks "renormalize"
to Gaussian ones. Yet, a main difference between
a simple random walk and the BS model is the presence
of memory effects. Indeed, any time a site is chosen (either as the
active one or as one of its neighbors), any memory
of its previous updates is lost. Therefore, the interaction
between the geometric structure given by the choice of the neighbors
according to $P(x)$ and the updates of the corresponding variables
can give rise to non-trivial effects.

As already mentioned above, the geometrical 
fractal dimension $D_f$ of the avalanches
is another quantity of interest. 
It is possible to relate $D_f$ to the other
exponents of the model. Indeed, we recall that the fractal
dimension relates the volume $N$ of the avalanche (that is, the number of sites
touched by the avalanche) with the typical size $R$ of the avalanche, as
$N \sim R^{D_f}$. On the other hand, $N$ scales with the duration
of the avalanche as $N \sim t^\mu$. 
The relation between the typical size of the avalanche $R$, and its
duration $t$ is given by $t \sim R^z$, $z$ being the dynamical exponent of 
the model. Then we find $D_f = z \mu$.

In principle, in order to determine $z$ it is possible to use
the all return time distribution $P_a(t)$. At long times 
on a $d-$dimensional lattice of $N=L^d$ sites, $P_a(t)$ flattens. W
e can write a scaling
form for $P_a(t)$ as
\begin{equation}
P_a(t,L) = t^{-\tau_a} f\left(\frac{t}{L^z}\right)
\label{scaling form}
\end{equation}
with $f(x)\sim const$ when $x\to 0$ and $f(x) \sim x^{\tau_a}$ when
$x \to \infty$. In this second case we find that
$P_a(t,L) = L^{-d}$ (roughly speaking,
as soon as every site of the lattice has been touched by 
an avalanche, they have all the same
probability to be chosen), from which we have $z=d/\tau_a$. 

Then we can  write an expression for the fractal dimension
\begin{equation}
D_f = d \frac{\mu}{\tau_a}\;\;\;.
\label{fractal d}
\end{equation}
This expression holds for the high dimensional simulations
of \cite{DMV98}: when avalanches are compact objects ($d=1,2$),
$\mu = \tau_a$. Then, $\mu \ne \tau_a$ and avalanches become
fractal objects. Indeed (\ref{fractal d}) approximates well
the data given in \cite{DMV98} for $d=3$ ($\tau_a=0.92$,
$\mu=0.85$, $D_f=2.6 \simeq 3 \mu/\tau_a = 2.77...$) and for $d=4$
($\tau_a=1.15$,
$\mu=0.92$, $D_f=3.3 \simeq 4 \mu/\tau_a = 3.2$).
As a byproduct, we find that (\ref{fractal d}) suggests an upper critical
dimension for the Bak-Sneppen model $d_{uc} = 6$: indeed
with $\mu=1$ and $\tau=3/2$ (the "mean-field", RNN, values of the
exponents) we find $D_f(d=6) = 4$, that is indeed the predicted
avalanche fractal dimension in the RNN limit.
$d_{uc} = 6$ is at odds with what stated in \cite{DMV98},
where $d_{uc} = 8$ was suggested by numerical simulations, but also with 
\cite{PMB96} where $d_{uc} = 4$ was claimed based on analogies  with
directed percolation.

\begin{figure}
\centerline{\psfig{file=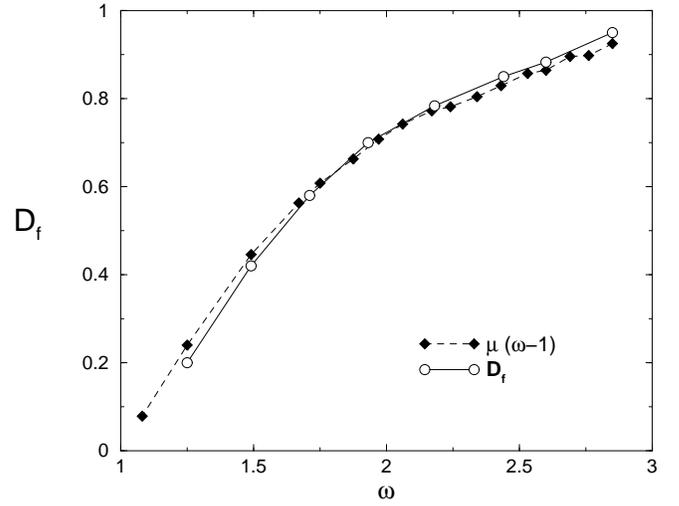,width=8.5cm,angle=0}}
\caption{Avalanche fractal dimension $D_f$ for different
values of $\omega$ (circles) and as from $D_f = (\omega-1) \mu$ (squares).}
\label{Fig: fig3}
\end{figure}

Although (\ref{fractal d}) seems to hold in the high dimensional case,
it does not fit the numerical results in the present LRNN approach.
The reason is that, whereas in the high dimensional 
case there is a single relation between time and space, namely
$t \sim L^{d/\tau_a}$, in the LRNN case there is a further relation,
the usual Levy random walk law $t \sim R^{\omega -1}$.
We measure the fractal dimension of avalanches using the
distance between the rightmost and leftmost touched sites as
a measure of $R$, and this corresponds to $z= \omega-1$.
Indeed, as it can be seen from Fig.\ref{Fig: fig3}, 
$D_f = \mu (\omega-1)$ approximates very well the measured fractal
dimensions.

As noted above about the fractal dimension of high-dimensional avalanches, 
we observed that $\mu=\tau_a$ corresponds to
compact avalanches. In the LRNN case we see from Fig.\ref{Fig: fig2} that
indeed $\mu \ne \tau_a$ for $\omega < 2$, even if  the 
fractal dimension $D_f<1$ already for $\omega<3$. We can try to 
understand this result remembering that $D_f$ is related to the 
random walk exponent $z=\omega-1$: although $z$ is 
different from its Gaussian value $z=2$ as soon as $\omega<3$, a 
Levy random walk with $\omega > 2$ is still compact, and so is the structure
built by a choice of neighbors according to (\ref{nn law}). 
Only when  $\omega < 2$ such a structure 
becomes genuinely fractal, and $\mu \ne \tau_a$.

\begin{figure}
\centerline{\psfig{file=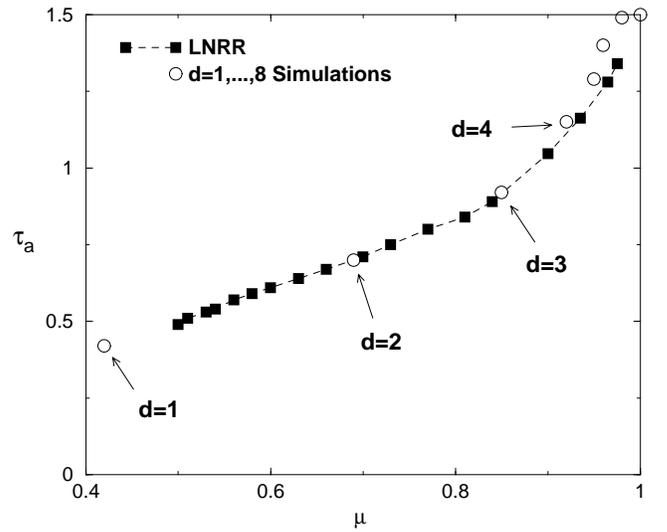,width=8.5cm,angle=0}}
\caption{All return time exponent $\tau_a$ vs. $\mu$ for different
values of $\omega$ (squares) and for different dimensions, from $d=1$
to $d=8$ (circles).}
\label{Fig: fig4}
\end{figure}

The possibility to obtain the exponent $D_f$ from $\mu$ and $\tau_a$ in 
high dimensions, and from $\mu$ and $\omega$ in the LRNN version,
suggests that indeed there are at most two independent exponents 
in the model, namely $\mu$ and $\tau_a$.
The LRNN model suggests the presence of a further 
(although non trivial) exponent relation.
Indeed, in Fig.\ref{Fig: fig4} we show the values of the 
$\tau_a$ exponent as a function of the
corresponding $\mu$ exponent for different values of 
$\omega$ and for different dimensions. 
As it can be seen, the agreement is good, suggesting that 
the knowledge of $\mu$ (or of $\tau_a$) is sufficient to 
know all the other exponents through
relations that, as for the $\mu/\tau$ one, could be highly non-trivial.

In conclusion, we have introduced a modification of the Bak-Sneppen model where
the neighbors of the active site are chosen at random over the lattice
with a probability that decreases like a power-law of 
the distance from the active site,
with an exponent $\omega$. As a result we find that 
the characteristic exponents
of the model interpolate between the $d=1$ limit 
($\omega = \infty$)
and the mean field (RNN) limit ($\omega \le 1$).
In particular, we verify that the known exponent 
relations hold for this model too.
Moreover, we find and verify an exponent relation for the 
fractal dimension of the avalanches, $D_f = (\omega -1) \mu$.
As a byproduct we obtain a relation between $D_f$ and 
$\mu$ and $\tau_a$ also in high dimensions,
fitting well the present numerical results up 
to $d=4$\cite{DMV98}, although it
suggests an upper critical dimension $d_{uc} = 6$ 
(and not $d_{uc}=4$ or $d_{uc}=8$
as previously believed). More accurate numerical simulations in high
dimensions are therefore needed. 
The relevance of the results reported in this 
Communication is manifold: they can
be looked at as an interesting modification of the Bak-Sneppen model, but their
full importance emerges when compared to the high
dimensional results of \cite{DMV98}. Indeed, they lead us to propose a new 
value of the upper critical dimension of the model, namely $d_{uc} =6$, and
to conjecture the existence of a still undiscovered 
exponent relation between $\mu$ and
$\tau_a$, reducing therefore the number of independent 
exponents to one in any dimension.
This results show therefore that there is still some way to go before a full
and satisfying understanding of the BS model is achieved.

The authors thank F. Slanina for useful discussions.
R. Cafiero and P. De Los Rios aknowledge financial 
support under the European network project FMRXCT980183.

\end{document}